\documentclass{moriond}

\def\Journal#1#2#3#4{{#1} {\bf #2}, #3 (#4)}

\def\PRL{\em Phys. Rev. Lett.}
\def\PRD{{\em Phys. Rev.} D}

\def\be{\begin{equation}}
\def\ee{\end{equation}}
\def\bea{\begin{eqnarray}}
\def\eea{\end{eqnarray}}

\newcommand{\Photo}{}

\begin{document}

\begin{flushright}
INR-TH-2016-013
\end{flushright}

\vspace*{3.0cm}

\title{Cosmological constraints on Lorentz Invariance violation \\
in gravity and dark matter }

\author{ Mikhail M. Ivanov}

\address{
FSB/IPHYS/LPPC, \'Ecole Polytechnique F\'ed\'erale de Lausanne, Lausanne, Switzerland\\
Faculty of Physics, Moscow State University, Moscow, Russia\\
Institute for Nuclear Research of the
Russian Academy of Sciences,
Moscow, Russia}

\maketitle\abstracts{
This brief contribution is devoted to phenomenological consequences of 
deviations from  Lorentz invariance in gravity and dark matter. 
We will discuss main effects on cosmological observables and 
current constraints derived from astrophysical and cosmological data.
}

\section{Introduction}

Lorentz invariance (LI) has been playing a key role in our understanding of gravity over many decades.
LI is a cornerstone of General Relativity (GR), the gravity theory whose predictions
were verified at various scales. 
Besides, deviations from LI were tightly constrained in the sector of Standard Model (SM) particles\cite{Liberati:2013xla}.
Yet there are reasons to place LI under scrutiny.
First, the constraints on Lorentz Invariance violation (LV) are much weaker or even do not exist for other sectors, e.g. the Dark sector. 
Second, theories without Lorentz invariance can have much better quantum behaviour 
than GR (e.g. Ho{\v{r}}ava gravity \cite{Barvinsky:2015kil,Horava:2009uw}).
Furthermore, abandoning LI leads to curios theoretical models which may 
address the mysteries of dark matter, dark energy, and inflation.
These reasons naturally pose the question: to what extent observational data support the belief that 
LI is a fundamental symmetry of Nature valid in all sectors? 
In this short note we will outline how this symmetry can be tested with cosmology.

\section{Gravity theory without Lorentz invariance}

The common way to parametrize generic effects of LV in gravity 
is to introduce a unit time-like vector field $u_\mu$ ("aether")
that breaks local LI of GR down to
the subgroup of rotations. 
This program yields the Einstein-aether
theory\cite{Jacobson},
\begin{eqnarray}
\label{khronoact}
&S_{[EH]}+S_{[u]} &=-\frac{1}{16\pi G_0}\int d^4x \sqrt{-g}
\bigg[R+K^{\mu\nu}_{\phantom{\mu\nu}\sigma\rho}\nabla_\mu u^\sigma\nabla_\nu u^\rho+l(u_\mu u^\mu-1)\bigg]\;,\\
&K^{\mu\nu}_{\phantom{\mu\nu}\sigma\rho}&=
c_1g^{\mu\nu}g_{\sigma\rho}
+c_2\delta^\mu_\sigma\delta^\nu_\rho
+c_3\delta^\mu_\rho\delta^\nu_\sigma
+c_4 u^\mu u^\nu g_{\sigma\rho}\,.
\end{eqnarray}
The dimensionless parameters $c_i$ characterize the deviation from GR and will be referred to as LV parameters. 
The scale $G_0$ controls the
gravitational interaction, and is related to the Newton's gravitational constant through the LV parameters.
Requiring that Minkowski space be a
background with stable perturbations implies certain inequalities for the LV parameters\cite{Blas:2014aca}.
Another set	of constraints comes from the absence of gravitational Cherenkov radiation\cite{Moore:2001bv}.
A variant of Einstein-aether gravity can be obtained by restricting the aether to be hypersurface-orthogonal, i.e.,
\begin{equation}
\label{khrono}
u_\mu\equiv {\partial_\mu \varphi}/{\sqrt{\partial^\nu \varphi \partial_\nu \varphi}}\,.
\end{equation}
In this case the action (\ref{khronoact}) corresponds to the so-called khronometric model \cite{Blas:2009qj}, 
the low-energy limit of Ho{\v{r}}ava gravity \cite{Horava:2009uw}. The restriction (\ref{khrono})
implies that $c_i$ are not independent, and the number of free LV parameters can be reduced to just three,
\begin{equation}
\label{params}
\lambda\equiv c_2\,,\quad \beta=c_3+c_1\,,\quad \alpha=c_1+c_4\,.
\end{equation}
In fact, Einstein-aether gravity and khronometric gravity have completely similar tensor and scalar sectors
which are characterised by these three parameters. The difference between the two theories appears only at the
level of vector perturbations, whose impact, however, is negligible 
in cosmology for most of the parameter space  \cite{ArmendarizPicon:2010rs}. 
Thus, for our purposes the Einstein-aether and khronometric gravity are almost identical and will be discussed in parallel in what follows.

The rest of fields in the Universe may also couple to $u_\mu$ through all possible covariant operators. 
For the SM sector, 
one has strong constraints that make this coupling negligible for any astrophysical observable \cite{Liberati:2013xla}.
Regarding dark matter (DM), the effect of a vector $u^\mu$ on a pressureless perfect fluid can be summarized in the following {\it macroscopic} 
action\cite{Blas:2012vn}
\begin{equation}
\label{eq:uDM}
S_{[DM-u]}=m\int d^4x\sqrt{-g} \,n \,F(v_\mu u^{\mu})\,,
\end{equation}
where $m$ is the mass of  DM particles, $n$ is their number density, and $v_{\mu}$ is their four-velocity. 
The function $F(v_\mu u^{\mu})$ parameterizes the interaction between the DM fluid and $u^\mu$. 
One can show that all the effects of LV in DM are encapsulated by a single parameter $Y$.
It is instructive to write down 
the energy-momentum tensor generated by the action (\ref{eq:uDM}),
\begin{equation}
T^{[DM-u]}_{\mu \nu}=\rho_{[DM]} v_\mu v_\nu-Y\rho_{[DM]}(u^\lambda v_{\lambda})v_\mu v_\nu+Y\rho_{[DM]}(u^\lambda v_{\lambda}) u_\mu u_\nu\,.
\end{equation}

One can also consider the coupling of the aether to the dark energy\cite{Blas:2011en}
or inflaton\cite{ArmendarizPicon:2010rs,Ivanov:2014yla} sectors. These interesting scenarios, however, are beyond the scope of this note.
In what follows, we assume that the recent accelerated expansion of the universe is driven by the usual cosmological constant.

\section{Astrophysical bounds}

Even though the aether does not couple directly to SM, it alters the gravitational interactions in the Solar system.
These effects are summarized in two PPN parameters
$\alpha_1^{PPN}$ and $\alpha_2^{PPN}$ depending on $\alpha$, $\beta$ and $\lambda$ \cite{Blas:2014aca}. 
These parameters are bounded as 
$|\alpha_2^{PPN}|<10^{-7}$ and $|\alpha_1^{PPN}|<10^{-4}$ \cite{Will:2005va}, which impose
two conditions of the same order of magnitude on the LV parameters. 
These bounds are very strong and, if imposed, exclude any observable consequences in cosmology.
However, these bounds can be avoided for a special choice of parameters, which is different in khronometric and Einatein-aether gravity\footnote{This difference arises due to vector polarisations, which contribute to the PPN parameters.},
\begin{eqnarray}
\alpha &=& 2\beta \,,\quad \quad \quad  \,\,\, \textit{khronometric}\,,\\
\label{eappn}
\alpha &=& -\beta-3\lambda \,,\quad \textit{Einstein-aether}\,,
\end{eqnarray}
since then $\alpha_1^{PPN}=\alpha_2^{PPN}=0$. 

In either case, the Solar system tests do not completely constrain (\ref{khronoact}). 
LV gravity modifies the dynamics of relativistic binaries by changing the orbits (strong PPN parameters) and the properties of gravitational radiation. 
Using data from binary pulsar systems one can constrain the LV parameters at the level\cite{Yagi:2013ava}
\begin{equation}
\label{eq:order}
|\alpha, \beta,\lambda|< 0.01 \,. 
\end{equation}

The parameter $\beta$ controlling the propagation of gravity waves has recently been constrained from the direct observation of the GW150914 event\cite{Blas:2016qmn},
\begin{equation} 
\beta<0.6\,.
\end{equation}

\section{Effects on cosmology and observational bounds}

It is customary to assume
that the preferred frame aligns at late times with the cosmic microwave background (CMB) frame. 
This is a suitable assumption because this situation is dynamically stable in case of misalignment between the two frames \cite{Carruthers:2010ii}. 
At the homogeneous and isotropic level, the only effect of LV on cosmology is 
a difference between the gravitational constants appearing in 
Friedman and Poisson equations,
\begin{equation}
\label{eq:G_G}
\Upsilon\equiv G_{Poisson}/G_{Friedman}-1=(\alpha+\beta+3\lambda)/{2}+O(\alpha^2)\,.
\end{equation}
  This is true even if LV in DM (\ref{eq:uDM}) is considered. 
Phenomenological consequences are richer once one considers cosmological perturbations\cite{Audren:2014hza,Audren:2013dwa}.  
One can identify the following effects that modify the power spectra of
CMB temperature fluctuations and {\it linear} perturbations of matter,

1. {\bf Enhanced gravity.} 
Thus effect is caused by the difference between $G_{Friedman}$ and $G_{Poisson}$ and
it modifies the gravitational interaction for {\it all}$\;$ species entering the Poisson equation. 
In particular, it leads to the enhanced growth rate of matter perturbations with respect 
to $\Lambda$CDM,
\begin{equation}
\label{enh}
\delta_{[DM]}\sim \delta_{[B]}\sim a(\tau)^{1+\frac{3}{5}\Upsilon}\,,
\end{equation}
where $a(\tau)$ is the scale factor.
For CMB, 
this effect boosts the Integrated Sachs-Wolfe contribution (ISW), 
and changes the amplitude and the phase of acoustic oscillations. 
This induces the enhancement of the CMB power spectrum at large scales and modifies the pattern of acoustic peaks.
The accelerated growth of perturbations (\ref{enh}) also enhances the amplitude of the matter power spectrum (MPS)
and changes its slope.
The parameter $\Upsilon$ accidentally vanishes 
in Einstein-aether gravity
once the PPN bound (\ref{eappn}) is imposed. Thus, the effect of enhanced gravity is viable only in the khronometric theory.

2. {\bf Gravitational slip.} 
Another generic effect of LV in gravity is the appearance of an extra contribution to the anisotropic stress,
which produces the difference between two gravitational potentials in the Newton gauge.
This increases the viscosity of cosmic fluid and damps out perturbations of all species.
The effect on CMB and MPS is an overall suppression, which, however, 
can be compensated by rescaling the amplitude of initial perturbations.
This degeneracy implies that  
gravitational slip should be less constrained from the data.

3. {\bf Scale-dependent enhancement of dark matter clustering.}
This effect is related to LV in dark matter and arises due to the violation of the equivalence principle in this sector.
Remarkably, there appears an analogue of the chameleon mechanism in this case.
The modes shorter than the 
characteristic "screening" scale $k_Y\sim H_0\sqrt{Y/a(\tau)\alpha}$ ($H_0$ is the Hubble rate today)
experience the accelerated growth of density fluctuations, 
\begin{equation}
\delta_{[DM]}\sim a(\tau)^{1+\kappa}\,, \quad \kappa=\frac{3}{5}\frac{\Omega_{dm}}{\Omega_{dm}+\Omega_{b}}Y+O(Y^2)\,,
\end{equation}
while for the modes larger
than $k_Y$ the effect of LV in DM 
is screened and they evolve as in $\Lambda$CDM.
This results in boosted ISW in CMB, and the scale-dependent enhancement of the matter power spectrum.
We emphasize that, even if the effects of  $Y\neq 0$ can be large, they may be completely screened at linear scales.

The discussed above effects were constrained using a combination of LSS observations and CMB data 
\cite{Audren:2014hza}.
The constraints on the LV parameters in gravity and dark matter 
appeared to be very similar
for Einstein-aether and khronometric gravity, and are given by (at $95\%$ CL),
\begin{eqnarray}
&\alpha &< 1.1\times 10^{-3}\,, \quad {(\beta+\lambda)}/{\alpha}<55\,,\,\,\, \quad Y<0.029\,, \quad \;\;\;\textit{khronometric}\,,\\
&\alpha &< 5.0\times 10^{-3}\,, \quad {(\beta+\lambda)}/{\alpha}<240\,, \quad Y<0.028\,, \quad \;\;\;\textit{Einstein-aether}\,.
\end{eqnarray} 

If one assumes LI in the sector of DM ($Y\equiv 0$), then the constraints are quite different for the 
Einstein-aether and khronometric 
models\cite{Audren:2014hza,Audren:2013dwa,Frusciante:2015maa}. 
For the khronometric theory one obtains, essentially, the same constraints on LV parameters in gravity as above.
In the Einstein-aether model the effect of enhanced gravity disappears and we are left only with gravitational slip which 
is partially degenerate with the amplitude of the primordial power spectrum. 
This makes the bounds on Einstein-aether gravity degrade by one order of magnitude
with respect to khronometric gravity, 
\begin{eqnarray}
&\alpha &< 1.8\times 10^{-3}\,, \quad {(\beta+\lambda)}/{\alpha}<91\,,\,\,\,  \quad \;\;\;\textit{khronometric}\,,\\
&\alpha &< 1.0\times 10^{-2}\,, \quad {(\beta+\lambda)}/{\alpha}<430\,,  \quad \;\;\;\textit{Einstein-aether}\,.
\end{eqnarray} 

We point out that so far all cosmological constraints were derived using the CMB and linear MPS data 
(in fact, the bounds presented above are saturated already with the CMB data).
We expect to obtain much better bounds once the non-linear dynamics at short scales (less than $10$ Mpc) 
is taken into account, e.g. along the lines of \cite{Blas:2015qsi}.

LV in DM can also be constrained from observations of virialized objects, such as DM halos\cite{Blas:2012vn}.
In fact, in addition to the violation of the equivalence principle, LV in DM generates velocity-dependent interactions,
which are negligible in cosmology but might affect the dynamics of galaxies and galaxy clusters. These interactions can be relevant only 
for fast ($v^2_{h}>\alpha/Y$) and 
small enough DM halos $R_h <\sqrt{{Y\rho_{halo}G_0}/{\alpha}}$ ($\rho_{halo}$ is the halo density).
For halos with larger radii all the effects of LV are eliminated by the screening mechanism.

To sum up, theories without Lorentz invariance provide us with an interesting 
framework to test modified gravity models and better understand the role of the Lorentz symmetry.
These theories yield peculiar phenomenological effects, which can be examined with cosmological data.
The cosmological constraints obtained so far bound LV in gravity at the sub-percent level, and
are competitive with those derived from astrophysical test. The bounds on LV in DM were put at the percent level.
More studies are required in order to understand the non-linear effects that will allow to test
Lorentz invariance to a new level of precision with upcoming surveys. 

{\it Acknowledgements}. This work was supported 
by the Swiss National Science Foundation
and the RFBR grant 14-02-00894.

\end{document}